# Covariance spectroscopy of molecular gases using fs pulse bursts created by modulational instability in gas-filled hollow-core fiber

MALLIKA IRENE SURESH[1,2*], PHILIP ST.J. RUSSELL[1,2], AND FRANCESCO TANI[1]

[1]*Max Planck Institute for the Science of Light and* [2]*Department of Physics, Friedrich-Alexander-Universität, Staudtstrasse 2, 91058 Erlangen, Germany*
*\*mallika-irene.suresh@mpl.mpg.de*

**Abstract:** We present a technique that uses noisy broadband pulse bursts generated by modulational instability to probe nonlinear processes, including infrared-inactive Raman transitions, in molecular gases. These processes imprint correlations between different regions of the noisy spectrum, which can be detected by acquiring single shot spectra and calculating the Pearson correlation coefficient between the different frequency components. Numerical simulations verify the experimental measurements and are used to further understand the system and discuss methods to improve the signal strength and the spectral resolution of the technique.

## 1 Introduction

Low noise trains of ultrashort pulses are a powerful tool for studying the properties of matter. As the duration gets shorter and the carrier frequency higher, however, realizing sources with sufficient peak and average power becomes difficult. Additionally, the complexity of the experiments relying on such sources limits their use to specialized laboratories. Consequently, over the years methods based on stochastic light sources have been developed to provide a simpler alternative [1-10]. Such methods often have short acquisition times, are robust against perturbation and do not require precise control of dispersion. In the majority of these noise-aided techniques, information about the sample is extracted using higher-order moments rather than the mean of repeated measurements acquired using low noise sources. Noise-aided techniques are useful in a variety of applications, from spatial and temporal imaging [9-11] to spectroscopy [1-3,6-8,12]. For instance, free electron lasers (FELs) can deliver high energy fs pulses in the x-ray spectral region [4,5], but usually exhibit strong stochastic fluctuations, which permits their use in ghost imaging [13,14]—an absorption-based method that relies on simultaneous measurements at multiple points and that has been demonstrated with both classical and quantum sources in the spatial, temporal and spectral domains. Although nonlinear processes do not always lead to absorption, they do imprint spectral correlations via frequency conversion. Taking advantage of this, noisy light sources were used to study the nonlinear response of materials as early as the 1980s [1,2]. Temporally incoherent nanosecond pulses have been used to measure, for example, the frequency and dephasing times of Raman vibrational modes as well as the nonlinear optical Kerr response in different organic liquids with sub-picosecond resolution [1,2,15,16] – the temporal resolution set by the coherence time of the pulses and not their duration. More recently, multidimensional spectroscopy using incoherent light has been demonstrated, and also simple single-beam coherent anti-Stokes Raman scattering (CARS) spectroscopy has been reported with deliberately randomised femtosecond pulses [3,7]. The energy conversion between different frequencies due to such nonlinear processes can be tracked by correlation maps. These maps have been used in different fields such as quantum optics [17], mass spectroscopy [18,19], nuclear magnetic resonance spectroscopy [20], nonlinear fiber optics [21,22], and very recently, Tollerud *et al*. used them to characterise solid samples by stimulated Raman scattering (SRS) of broadband incoherent femtosecond pulses [6].

Here we report that a noisy supercontinuum (SC) generated by modulational instability (MI) can be used to measure the nonlinear response of a sample, in spite of large shot-to-shot variations in the spectral power distribution. In gas-filled hollow-core photonic crystal fiber (HC-PCF) pumped with µJ-level pulses, MI can generate a PHz-wide SC extending from the near-IR to the UV [23,24]. In the time domain, this SC consists of incoherent bursts of few fs (or shorter) pulses with sufficient peak power both to pump and probe a sample. Information about the sample can be extracted by acquiring an ensemble of random spectra and calculating the correlation between spectral components that are imprinted on the SC by the nonlinear response of the sample [6]. In this way, we probe infrared-inactive vibrational Raman transitions in molecular gases, and demonstrate scan-free nonlinear spectroscopy with a few seconds acquisition time. This could be further reduced to sub-millisecond timescales by employing a higher repetition rate laser source combined with spectral measurements by dispersive Fourier transformation.

## 2 Experiment and numerical simulations

The experimental set-up is shown in Fig. 1(a). A noisy supercontinuum was generated by pumping a single-ring PCF (30 µm core diameter, ~220 nm core wall thickness and 25 cm length, filled with 22 bar of Ar) with pulses from a Ti:sapphire laser (100 fs, 805 nm, 7.2 µJ, 1 kHz). This fiber is henceforth referred to as the source fiber. The resulting bursts of incoherent ultrashort pulses, generated by MI, were collimated with parabolic mirrors (15 cm focal length) and launched into a sample fiber which is a 7-mm-long kagomé-PCF (34 µm core diameter) placed in a second gas cell containing the sample molecular gas. The sample fiber was kept as short as possible so as to suppress the effects of weaker instantaneous nonlinearities, which introduce spectrally uniform frequency correlations resulting in a lower signal-to-noise ratio (SNR). The pulse energy in the sample fiber, though limited to ~1.8 µJ by loss in the optical components, was sufficient for the experiments. Single-shot spectra of the light emerging from the sample fiber were recorded using a triggered fiber-coupled spectrometer with a spectral resolution of ~1.5 nm (2.4 THz) at ~430 nm. The CCD detector was intensity calibrated, taking account of its nonlinear response by following the procedure described in [25]. A low-pass filter with cut-off frequency ~606 THz (495 nm), placed before the sample fiber, was used to filter out higher frequency signals in the supercontinuum, allowing anti-Stokes correlations to be detected against a dark background and thus improving the SNR.

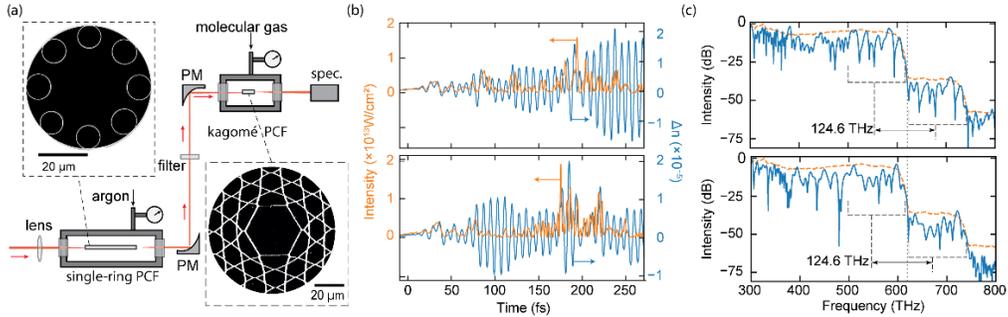

**Figure 1.** (a) Experimental set-up: the pulses undergo MI in the source fiber (left inset: 30 µm core single-ring PCF) and then are used to probe the gas in the sample fiber (right inset: 34 µm core kagomé PCF), parabolic mirrors (PM) are used to collimate and refocus the light. The filter removes frequencies above ~606 THz. (b) Two single-shot measurements of simulated noise bursts, showing their temporal profiles (orange, left axis) and the corresponding index change induced by Raman scattering $\Delta n$ (blue, right axis) in $H_2$. (c) Single-shot plots of the simulated spectra (blue), showing replication of spectral features separated by the Raman shift in $H_2$ (125 THz), which are not visible when averaging over many shots (orange dashed).

A unidirectional nonlinear wave equation was used in the numerical modeling [26,27], the Raman polarization term being approximated by the response function described in [28]. To model the experiment we simulated the propagation of a 5.8 µJ, 100 fs pulse first through the Ar-filled source fiber and then, after filtering out frequencies above 606 THz, through the sample fiber filled with $H_2$. Fig. 1(b) shows simulated single-shot temporal profiles of two pulse bursts (orange curves) sent to the sample fiber and the corresponding nonlinear refractive index change induced by stimulated Raman scattering $\Delta n$ in $H_2$ (blue curves). The envelope duration of the bursts (not shown in the plot) is ~300 fs (full width half maximum). Each of the temporal spikes in each incoherent burst is shorter than a half-cycle of molecular vibrations in hydrogen (8 fs), which has the largest Raman shift of all molecular gases, allowing impulsive excitation of Raman coherence in any molecular gas. Also, the spikes will be blue or red-shifted depending on the temporal slope of the refractive index wave (caused by the Raman coherence) upon which they are riding. If the index increases with time, the spike frequency will be red-shifted, and vice-versa. If the index slope is zero, the spike will maintain the same frequency while broadening or narrowing in bandwidth [29]. More intense peaks will contribute more strongly to the coherence, resulting in more photons being scattered to higher/lower frequencies. Conversely, lower intensity peaks will generate weaker coherence and fewer photons will be scattered. Thus, spectral features separated by the Raman frequency will be positively correlated in intensity, permitting them to be measured.

Since other correlations between the frequencies across the spectrum are destroyed by employing a series of random spectra, the Raman-induced correlations can be identified by calculating higher-order moments. Here, as in [6], we use the Pearson correlation coefficient which is defined for pairs of frequencies as:

$$\rho(\omega_i,\omega_j) = \frac{\langle I(\omega_i)I(\omega_j)\rangle - \langle I(\omega_i)\rangle\langle I(\omega_j)\rangle}{\sqrt{\left(\langle I^2(\omega_i)\rangle - \langle I(\omega_i)\rangle^2\right)\left(\langle I^2(\omega_j)\rangle - \langle I(\omega_j)\rangle^2\right)}}, \quad (1)$$

where $I(\omega_i)$ is a time-series array of intensities at optical angular frequency $\omega_i$ and the angular brackets denote the average calculated over the ensemble. These coefficients vary over the range $-1 \leq \rho(\omega_i, \omega_j) \leq 1$, $\rho(\omega_i, \omega_j)$ being positive when the intensities at the two frequencies are correlated, negative when they are anti-correlated and zero when they are completely uncorrelated. The resulting correlation map of $\rho(\omega_i, \omega_j)$ versus $\omega_i$ and $\omega_j$ (Fig. 2) is symmetric across the diagonal $\omega_i = \omega_j$, where the correlation is always +1.

It is interesting to draw an analogy with the work of Katz et al. [30], who shaped short coherent pulses with a notch filter and used them to impulsively excite and probe the vibrational Raman transitions of a sample. In that work the vibrational spectrum was obtained by detecting CARS spectra and measuring the blue-shift of narrow spectral features from the notch frequency. The spectroscopic resolution was therefore set by the spectral width of the notch filter. Similarly, in our work each burst of incoherent spikes has fine spectral features that generate CARS spectra modulated in the same way, blue-shifted by the molecular vibrational frequency [7,31] and evident in the simulated spectra in Fig. 1(c). Red-shifted spectra generated by coherent Stokes Raman scattering (CSRS) can also be detected by filtering out low frequencies from the SC while keeping its spectral bandwidth larger than the Raman shift to be detected. In both cases the spectroscopic resolution is set by the average width of the spectral features in the SC, which depends both on the soliton order of the pulses at the input of the source fiber and the temporal duration of the noise burst at the output of the source fiber.

## 3 Results

Figure 2(a) (upper) shows the mean (solid line) and standard (shaded area) deviations of 15,000 random spectra recorded at the output of the second fiber when it was evacuated (left) and filled with 15 bar $H_2$ (right). Below are the correlation maps obtained by calculating the Pearson coefficients of the ensemble of spectra. Although the spectra in each case are almost indistinguishable, in the hydrogen-filled sample fiber (second plot of the spectra) a weak optical signal can be seen just above the noise floor, in the region above the filter cut-off (~620 – 750 THz). This is the generated anti-Stokes signal. In the corresponding correlation map, it is visible as clear correlations between pump and anti-Stokes frequencies when the anti-Stokes frequency lies above the filter cut-off. These correspond to impulsively excited Raman coherence in the hydrogen molecules that upshifts the pump photons by $\Omega_R/2\pi = 125$ THz. The faint horizontal bar (vertical in the upper half of the plot) are caused by Kerr-related four-wave mixing that generates a side-band at the filter cut-off frequency.

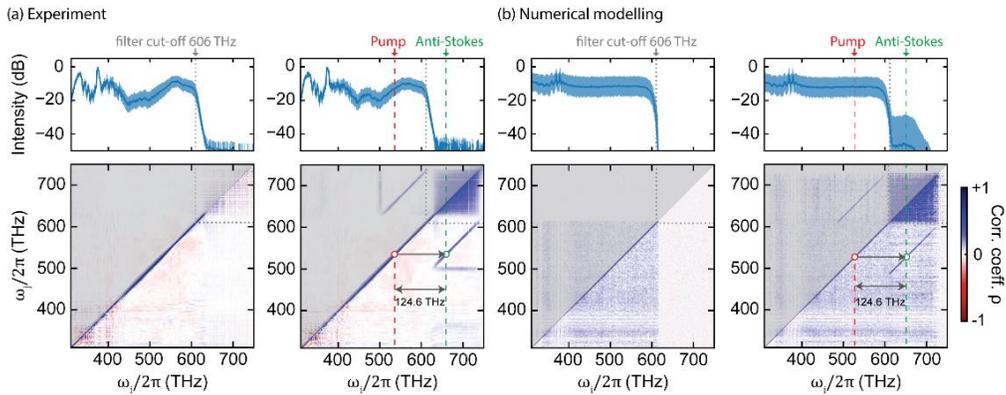

**Figure 2.** (a) Upper panes: measured spectra from 15,000 measurements in a 7-mm-long sample fiber. The solid line denotes the mean and the blue-shaded area the standard deviation. The gray dashed lines mark the filter cut-off. Left: evacuated. Right: filled with 15 bar $H_2$. Lower planes: correlation maps, which are symmetric about the diagonal (the upper half is grayed out). For illustration, one pump frequency is marked (red circle). It is correlated with an anti-Stokes signal (green circle) with a frequency difference of 125 THz. (b) Results of 900 simulations for the same parameters as in (a).

To model the experimental results we simulated propagation of a 100 fs pulse first through the Ar-filled source fiber and then, after spectral filtering, through the $H_2$-filled sample fiber. The results of the simulations, averaged over 900 noise-seeded realizations, show good agreement with experiment (Fig. 2(b)). Note that it is not necessary to simulate thousands of laser shots (this can take several hours) to find the optimal parameters. We found that the main outcome was unaffected when numerically generated pulses with random phase, amplitude and similar spectral bandwidth were used as input to the second fiber. Note also that dispersion plays only a marginal role in the determination of the Raman frequency shift and its linewidth and thus can be safely neglected, further reducing the computational time.

A CARS signal is generated when photons at frequency $\omega_p$ in the noisy pump spectrum are up-shifted to the anti-Stokes band at $\omega_p + \Omega_R$, resulting in correlations $\rho(\omega_p, \omega_p + \Omega_R) > 0$ between these pairs of frequencies, as seen in Fig. 2. For frequencies close to the low-pass filter, the correlation lines are markedly clearer because the filter removes high frequency light from the pump before it enters the gas, improving the SNR and resulting in clear anti-Stokes correlation signals for $\omega_i$ values between $\omega_f$ and $\omega_f + \Omega_R$, where $\omega_f$ is the cut-off frequency of the filter.

The correlation spectra in Fig. 3(a) correspond to mean correlation value taken across the extension of the correlation bands of maps with different gases (as shown in Fig. 2(b) for $H_2$ but not shown here for the other gases), plotted against the frequency separation $\delta\omega = \omega_i - \omega_j$ along the horizontal axis from the diagonal. In the limit of continuous $\omega$, this can be expressed mathematically as:

$$\rho_{mean}(\delta\omega) = \frac{1}{\Omega_R} \int_{\omega_f - \Omega_R}^{\omega_f} \rho(\omega_j + \delta\omega, \omega_j) d\omega_j, \qquad (2)$$

plotted against $\delta\omega$ over the range 15 THz $\leq \delta\omega/2\pi \leq$ 150 THz (i.e. the range spanned by the horizontal axis in Fig. 3(a)) and the limits of the integral are defined over the range $(\omega_f - \Omega_R) \leq \omega_j \leq \omega_f$, which is where the Raman-induced correlation band appears. For discrete values of $\omega$, this is simply the arithmetic mean. The evacuated sample fiber produces no correlations (purple curve marked "vacuum" in Fig. 3(a), corresponding to the data shown in Fig. 2(a)). The other curves correspond to the data obtained when the sample fiber was separately filled with $SF_6$, $N_2$, $CH_4$ and $H_2$ at 15 bar.

Although 15,000 single-shot spectra were collected for each gas, the SNR reached a maximum typically after only 5,000 shots. In the case of $CH_4$ the CARS spectra were typically much brighter than for the other gases, indeed a strong correlation signal could even be detected in free-space. One of the reasons for this is that the correlation signal depends not only on the gain of the Raman process, but also inversely on the dephasing time $T_2$ (at 15 bar, for $H_2$: $T_2 \sim 430$ ps and for $CH_4$: $T_2 \sim 23.6$ ps [32]), as can be seen from [15]. Each line-plot shows a clear correlation peak at a frequency corresponding to the vibrational Raman shift of the probed gas (indicated by the vertical dashed lines). We also observed a clear signal for $D_2$, but it is not included in the plot because its Raman shift is 89.7 THz, very close to the 87.6 THz of methane. Note that these vibrational Raman transitions are infrared-inactive [33], i.e., they do not appear in the infrared absorption spectrum and hence cannot be detected using absorption-based methods such as ghost-imaging in the spectral domain.

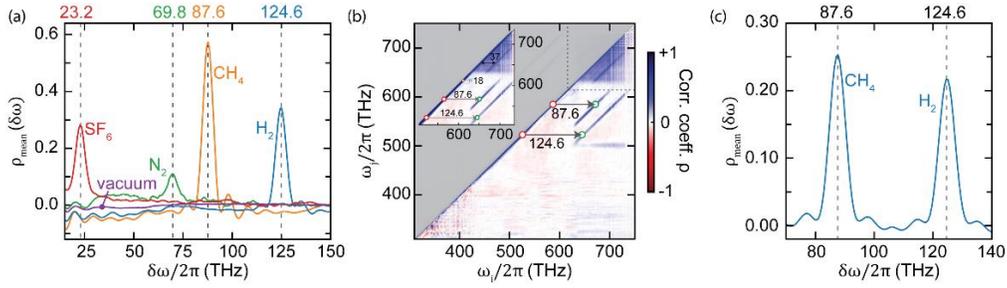

**Figure 3.** (a) Measured mean correlation coefficients plotted against frequency offset from the diagonal when the sample fiber is evacuated or filled with a single pure gas at 15 bar. The numbers above the plot are the Raman frequencies in THz for each gas species. (b) Correlation map calculated from spectra generated in a sample fiber filled with a mixture of 10 bar $CH_4$ and 10 bar $H_2$. Two illustrative correlations are marked in, one for each gas (124.6 THz for $H_2$ and 87.6 THz for $CH_4$). The inset shows a zoom-in showing the four off-diagonal lines: the two Raman transitions at 124.6 THz and 87.6 THz, 37 THz which is the frequency difference between them and the very faint rotational transition of $H_2$ at 18 THz. (c) The corresponding correlation spectrum for the gas mixture in (b).

Gas mixtures were also successfully probed, as seen in the correlation map in Fig. 3(b) for a sample fiber filled with 10 bar of $CH_4$ and 10 bar of $H_2$. Two anti-Stokes correlation lines, corresponding to the two vibrational shifts, are clearly visible. The faint lines at 37 THz, within the filter cut-off region, correspond to the frequency difference between the transitions of these two gases. Very weak signals from rotational Raman scattering in $H_2$ are also visible, in the vicinity of $\omega_i = \omega_j = 2\pi \times 600$ THz, offset by 18 THz from the diagonal. Although the rotational Raman signal is weaker than the vibrational [32], its visibility can be improved by further randomizing the spectral phase. This would serve to reduce both the background off-diagonal correlations and the width of the diagonal in the correlation map, thus making the correlation band more easily distinguishable. Furthermore, circularly polarised light can be used to enhance the gain of the rotational modes. A corresponding correlation spectrum with the two main peaks is shown in Fig. 3(c).

In both experiment and simulations we found that a clear correlation signal could be detected even when only 10 shots were acquired before calculating the map. By relaxing the need for triggered single-shot spectral acquisition, this would be of benefit in experiments using high repetition rate lasers.

## 4 Discussion

The advantage of using incoherent light with noise-aided techniques is that the temporal resolution is set by the correlation time of the pulses rather than their temporal duration [1] and hence, is determined by the bandwidth of the supercontinuum bursts used here. As the SC spectrum is a few hundred THz wide, the resulting coherence time is of order a fs (note however that if the spectral phase across the noise burst is not fully randomized, this coherence time will be slightly longer). Conversely, transform-limited pulses force one to trade temporal against spectral resolution.

However, with the noise-aided technique here, the spectral resolution is determined by the convolution of the spectrometer response with the width of the individual spectral features in the noisy bursts. These spectral features can be made very fine by allowing the few-cycle pulses within the noisy bursts to propagate over a longer source fiber, thereby increasing the nonlinear interaction and further randomizing the spectral phase of the noisy bursts [16]. In this manner, when a 60 cm source fiber is used in the first stage rather than a 25 cm one, the linewidth of the correlated signal for 6 bar of $CH_4$ (filled in the sample fiber) decreases from a full width half maximum (FWHM) of 6.7 THz to 5.3 THz. The linewidth can be further decreased to 4.1 THz if a filter with a 667 THz (450 nm) cut-off is used, since the high frequency side of the SC has finer features.

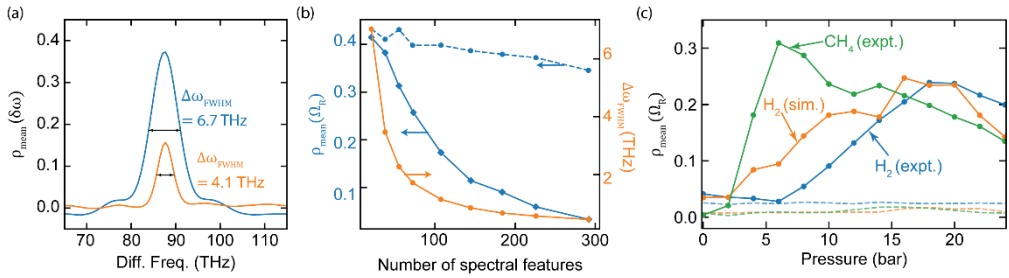

**Figure 4.** (a) Correlation signal for two different noise bursts measured for 6 bar of $CH_4$, showing a FWHM linewidth reduction from 6.7 to 4.1 THz, accompanied by a decrease in correlation strength from 0.38 to 0.15. (b) Simulated plot showing the decrease in signal strength (left axis) and linewidth (right axis) as the number of spectral features in the noise burst increases. The dashed line shows the result with an extended dynamic range of the spectral intensity. (c) Experimental (blue) and simulated (orange) values of the correlation coefficient as a function of $H_2$ pressure. The green curve shows experimental data for $CH_4$. The dashed lines show the corresponding background correlation signal for each case.

This improvement in the spectral resolution, however, comes at the cost of the strength of the correlation signal, which decreases by more than a factor of two, as seen in Fig. 4(a). This is also predicted by numerical simulations, as shown in Fig. 4(b). We define the number of identifiable spectral peaks in the SC (see, e.g., Fig. 1(c)) as the number of those that are at least 10% of the highest peak value (in the linear scale). This number may be regarded as a figure of merit for the spectral resolution, since it corresponds to the inverse spectral width of the SC features. Averaging numerically over 2,000 simulated shots with randomized phase and amplitude and neglecting the effects of dispersion, we obtain the plots in Fig. 4(b) of the strength $\rho_{mean}(\delta\omega=\Omega_R)$ and the linewidth $\Delta\omega_{FWHM}$ of the mean correlation coefficient. As the number of features increases, the correlation band becomes narrower (orange curve, right axis) and weaker (blue curve, left axis). The attenuation of the correlation signal is mainly due to the limited dynamic range of the spectrometer, as well as the much weaker CARS signal, whose spectral intensity is typically 30 dB below that of the SC. If a spectrometer with a greater dynamic range is used and the measured spectrum is attenuated at frequencies below the filter cut-off, the spectral resolution can be improved while maintaining a high SNR. The dashed blue curve in Fig. 4(b) plots the numerically modeled correlation strength over an extended dynamic range.

The coherence wave generated by pump-to-Stokes conversion has wavevector $\beta_{CW} = \beta(\omega_p) - \beta(\omega_p - \Omega_R)$, where $\beta(\omega)$ is the propagation constant (accurately given by the capillary approximation [34]). For strong CARS from a probe signal at a third frequency $\omega_{pr}$, $\beta_{CW}$ must be phase matched, i.e., $\Delta\beta = \beta(\omega_{pr}) - \beta(\omega_{pr} + \Omega_R) - \beta_{CW} = 0$. This may be achieved by tuning the gas pressure and composition, or by selecting a sample fiber with a different core diameter. The results of a series of measurements at different $H_2$ pressures are plotted in Fig. 4(c) (blue curve), for 60 cm of source fiber filled with 20 bar of argon, followed by a $\omega_f/2\pi$ ~667 THz (450 nm) cut-off filter before the second gas cell, which housed a 1-cm-long sample fiber. As can be seen, the correlation coefficient at the peak of $\rho_{mean}(\delta\omega)$, i.e. at $\delta\omega = \Omega_R$, reaches a maximum at 18 bar, the pressure at which the phase-mismatch is minimized. Carrying out the same measurement for $CH_4$ resulted in a signal that peaked at 6 bar, as shown by the green curve in Fig. 4(c). The corresponding curve for $H_2$ obtained from a simulated pressure scan is plotted as the orange curve in Fig. 4(c). For these experimental parameters, the correlation signal reduces to the background level at ambient pressure, as can be seen by the color-coded dashed lines, which indicate the background correlation signal in each case. With slightly higher energy in the SC and a shorter sample fiber, however, a correlation signal could be detected (at the cost of the linewidth) even at 1 bar of $CH_4$. As mentioned above, this could be improved by decreasing the difference in spectral intensity between the SC and the generated signal above the filter cut-off, and by optimizing phase-matching in the sample fiber.

It is interesting to note that although we have studied Raman scattering events which have dephasing times that are longer than the pulse duration, numerical simulations show that covariance-based spectroscopy could also be used to study dynamics with dephasing times shorter than the duration of the incoherent pulse burst, permitting liquid and solid samples to be probed as well.

## 5 Conclusions

An incoherent burst of ultrashort pulses generated by MI in a gas-filled HC-PCF can be used to probe infrared-inactive vibrational Raman modes of molecular gases using covariance spectroscopy. The technique requires only a single laser, is simple to implement and does not require scanning or pulse shaping. The generation of chaotic light (i.e. light with randomly varying shot-to-shot spectral phase) via MI is highly efficient (limited only by fiber loss), and the repetition rate can be scaled to the MHz range, permitting sub-millisecond acquisition times. Furthermore, the SC generated in gas-filled HC-PCF can be extended into the deep ultraviolet where it can be used for resonantly enhanced Raman spectroscopy. Interestingly, simulations suggest that covariance-based spectroscopy could also be used to study dynamics with dephasing times shorter than the duration of the incoherent pulse burst, permitting liquid and solid samples to be probed. In the experiments the spectral resolution was limited to ~4 THz. State-of-the-art studies using broadband single-beam CARS spectroscopy report sub-THz (20-30 cm$^{-1}$) spectral resolution [35, 36]. Numerical simulations of the technique reported here suggest that sub-THz resolution could be achieved by increasing the number of spectral features in the SC pulse burst. Finally, a spectral resolution exceeding that of the spectrometer may be achieved by simultaneously recording the spectrum before and after the sample and employing data analysis methods based on sparse-sampling [37].

## 6  Disclosures

The authors declare no conflicts of interest.

## 7  References

1. T. Hattori, A. Terasaki, and T. Kobayashi, "Coherent Stokes Raman scattering with incoherent light for vibrational-dephasing-time measurement," Phys. Rev. A **35**, 715–724 (1987).
2. M. J. Stimson, D. J. Ulness, and A. C. Albrecht, "Frequency and time resolved coherent Stokes Raman scattering in CS$_2$ using incoherent light," Chem. Phys. Lett. **263**, 185–190 (1996).
3. D. B. Turner, P. C. Arpin, S. D. McClure, D. J. Ulness, and G. D. Scholes, "Coherent multidimensional optical spectra measured using incoherent light," Nat. Commun. **4**, 2298 (2013).
4. V. Kimberg and N. Rohringer, "Stochastic stimulated electronic x-ray Raman spectroscopy," Struct. Dyn. **3**, 034101 (2016).
5. K. Meyer, C. Ott, P. Raith, A. Kaldun, Y. Jiang, A. Senftleben, M. Kurka, R. Moshammer, J. Ullrich, and T. Pfeifer, "Noisy Optical Pulses Enhance the Temporal Resolution of Pump-Probe Spectroscopy," Phys. Rev. Lett. **108**, 098302 (2012).
6. J. O. Tollerud, G. Sparapassi, A. Montanaro, S. Asban, F. Glerean, F. Giusti, A. Marciniak, G. Kourousias, F. Billè, F. Cilento, S. Mukamel, and D. Fausti, "Femtosecond covariance spectroscopy," Proc. Natl. Acad. Sci. **116**, 5383–5386 (2019).
7. X. G. Xu, S. O. Konorov, J. W. Hepburn, and V. Milner, "Noise autocorrelation spectroscopy with coherent Raman scattering," Nat. Phys. **4**, 125–129 (2008).
8. B. Dayan, A. Pe'er, A. A. Friesem, and Y. Silberberg, "Two Photon Absorption and Coherent Control with Broadband Down-Converted Light," Phys. Rev. Lett. **93**, 023005 (2004).
9. P.-A. Moreau, E. Toninelli, T. Gregory, and M. J. Padgett, "Ghost Imaging Using Optical Correlations," Laser Photonics Rev. **12**, 1700143 (2018).
10. O. Katz, P. Heidmann, M. Fink, and S. Gigan, "Non-invasive single-shot imaging through scattering layers and around corners via speckle correlations," Nat. Photonics **8**, 784–790 (2014).
11. P. Ryczkowski, M. Barbier, A. T. Friberg, J. M. Dudley, and G. Genty, "Ghost imaging in the time domain," Nat. Photonics **10**, 167–170 (2016).

12. C. Amiot, P. Ryczkowski, A. T. Friberg, J. M. Dudley, and G. Genty, "Supercontinuum spectral-domain ghost imaging," Opt. Lett. **43**, 5025–5028 (2018).
13. D. Ratner, J. P. Cryan, T. J. Lane, S. Li, and G. Stupakov, "Pump-Probe Ghost Imaging with SASE FELs," Phys. Rev. X **9**, 011045 (2019).
14. T. Driver, S. Li, E. G. Champenois, J. Duris, D. Ratner, T. J. Lane, P. Rosenberger, A. Al-Haddad, V. Averbukh, T. Barnard, N. Berrah, C. Bostedt, P. H. Bucksbaum, R. Coffee, L. F. DiMauro, L. Fang, D. Garratt, A. Gatton, Z. Guo, G. Hartmann, D. Haxton, W. Helml, Z. Huang, A. LaForge, A. Kamalov, M. F. Kling, J. Knurr, M.-F. Lin, A. A. Lutman, J. P. MacArthur, J. P. Marangos, M. Nantel, A. Natan, R. Obaid, J. T. O'Neal, N. H. Shivaram, A. Schori, P. Walter, A. L. Wang, T. J. A. Wolf, A. Marinelli, and J. P. Cryan, "Attosecond transient absorption spooktroscopy: a ghost imaging approach to ultrafast absorption spectroscopy," Phys. Chem. Chem. Phys. **22**, 2704–2712 (2020).
15. T. Kobayashi, A. Terasaki, T. Hattori, and K. Kurokawa, "The application of incoherent light for the study of femtosecond-picosecond relaxation in condensed phase," Appl. Phys. B **47**, 107–125 (1988).
16. T. Kobayashi, A. Terasaki, T. Hattori, and K. Kurokawa, "Femtosecond relaxation processes in nonlinear materials studied with incoherent light," Rev. Phys. Appl. **22**, 1773-1785 (1987).
17. Y. Zhang, D. England, A. Nomerotski, P. Svihra, S. Ferrante, P. Hockett, and B. Sussman, "Multidimensional quantum-enhanced target detection via spectrotemporal-correlation measurements," Phys. Rev. A **101**, 053808 (2020).
18. M. R. Bruce, L. Mi, C. R. Sporleder, and R. A. Bonham, "Covariance mapping mass spectroscopy using a pulsed electron ionizing source: application to $CF_4$," J. Phys. B: At. Mol. Opt. Phys. **27** 577–5794, (1994).
19. C. S. Slater, S. Blake, M. Brouard, A. Lauer, C. Vallance, J. J. John, R. Turchetta, A. Nomerotski, L. Christensen, J. H. Nielsen, M. P. Johansson, and H. Stapelfeldt, "Covariance imaging experiments using a pixel-imaging mass-spectrometry camera," Phys. Rev. A **89** 011401, (2014).
20. F. Bruschweiler and F. Zhang, "Covariance nuclear magnetic resonance spectroscopy," J. Chem. Phys. **120,** 5253 (2004).
21. B. Wetzel, A. Stefani, L. Larger, P. A. Lacourt, J. M. Merolla, T. Sylvestre, A. Kudlinski, A. Mussot, G. Genty, F. Dias, and J. M. Dudley, "Real-time full bandwidth measurement of spectral noise in supercontinuum generation", Scientific Report **2**, (2012).
22. P. Robert, C. Fourcade-Dutin, R. Dauliat, R. Jamier, H. Muñoz-Marco, P. Pérez-Millán, J. M. Dudley, P. Roy, H. Maillotte, and D. Bigourd, "Spectral correlation of four-wave mixing generated in a photonic crystal fiber pumped by a chirped pulse," Opt. Lett. **15**, 4148–4151 (2020).
23. F. Tani, J. C. Travers, and P. St.J. Russell, "PHz-wide Supercontinua of Nondispersing Subcycle Pulses Generated by Extreme Modulational Instability," Phys. Rev. Lett. **111**, 033902 (2013).
24. P. St. J. Russell, P. Hölzer, W. Chang, A. Abdolvand, and J. C. Travers, "Hollow-core photonic crystal fibres for gas-based nonlinear optics,"Nat. Phot. **8**, 278-286 (2014).
25. M. Nehir, C. Frank, S. Aßmann, E. P. Achterberg, "Improving Optical Measurements: Non-Linearity Compensation of Compact Charge-Coupled Device (CCD) Spectrometers," Sensors **19**, 2833 (2019).
26. M. Kolesik, J. V. Moloney, and M. Mlejnek, "Unidirectional Optical Pulse Propagation Equation," Phys. Rev. Lett. **89**, 283902 (2002).
27. F. Tani, J. C. Travers, and P. St.J. Russell, "Multimode ultrafast nonlinear optics in optical waveguides: numerical modeling and experiments in kagomé photonic-crystal fiber," J. Opt. Soc. Am. B **31**, 311-320, (2014).
28. F. Belli, "Ultrafast Raman scattering in gas-filled hollow-core fibers," Ph.D. thesis, (2017).

29. F. Belli, A. Abdolvand, J. C. Travers, and P. St.J. Russell, "Control of ultrafast pulses in a hydrogen-filled hollow-core photonic-crystal fiber by Raman coherence," Phys. Rev. A **97**, 013814 (2018).
30. O. Katz, J. M. Levitt, E. Grinvald, and Y. Silberberg, "Single-beam coherent Raman spectroscopy and microscopy via spectral notch shaping," Opt. Express **18**, 22693–22701 (2010).
31. S. T. Bauerschmidt, D. Novoa, B. M. Trabold, A. Abdolvand, and P. St. J. Russell, "Supercontinuum up-conversion via molecular modulation in gas-filled hollow-core PCF," Opt. Express **22**, 20566–20573 (2014).
32. M. J. Weber, "Handbook of Laser Science and Technology Supplement 2: Optical Materials," CRC Press (1994).
33. D. A. Long, "The Raman effect: A Unified Theory of Raman Scattering by Molecules" John Wiley & Sons Ltd. (2002).
34. J. C. Travers, W. Chang, J. Nold, N. Y. Joly, and P. St.J. Russell, "Ultrafast nonlinear optics in gas-filled hollow-core photonic crystal fibers," J. Opt. Soc. Am. B **28**, A11–A26 (2011).
35. O. Katz, A. Natan, Y. Silberberg, and S. Rosenwaks, "Standoff detection of trace amounts of solids by nonlinear Raman spectroscopy using shaped femtosecond pulses," Appl. Phys. Lett. **92**, 171116 (2008).
36. N. Dudovich, D. Oron, and Y. Silberberg, "Single-pulse coherently controlled nonlinear Raman spectroscopy and microscopy," Nature **418**, 512–514 (2002).
37. A. Boschetti, A. Taschin, P. Bartolini, A. K. Tiwari, L. Pattelli, R. Torre, and D. S. Wiersma, "Spectral super-resolution spectroscopy using a random laser," Nat. Photonics **14**, 177–182 (2020).